\documentclass[runningheads]{llncs}

\usepackage[T1]{fontenc}
\usepackage{hyperref}

\usepackage{float}
\usepackage{bbding}

\usepackage{booktabs} 
\usepackage{siunitx}
\usepackage{tcolorbox}
\tcbuselibrary{skins} 
\tcbuselibrary{breakable} 

\usepackage{booktabs}
\usepackage{siunitx}

\usepackage{xcolor}
\usepackage{graphicx}
\begin{document}
\title{An Empirical Study of Security-Policy Related Issues in Open Source Projects}
\titlerunning{Security-Policy Related Issues in Open Source Projects}
%
\author{
Rintaro Kanaji\inst{1}\Envelope{} \and 
Brittany Reid\inst{1} \and 
Yutaro Kashiwa\inst{1} \and 
Raula Gaikovina Kula\inst{2}  \and 
Hajimu Iida\inst{1}
}
\authorrunning{R. Kanaji et al.}
%

\institute{
Nara Institute of Science and Technology, Japan
\and
The University of Osaka, Japan \\
\email{Corresponding Author\Envelope{}: kanaji.rintaro.kq8@naist.ac.jp}
}
\maketitle              
\begin{abstract}
GitHub recommends that projects adopt a \texttt{SECURITY.md} file that outlines vulnerability reporting procedures. However, the effectiveness and operational challenges of such files are not yet fully understood. This study aims to clarify the challenges that \texttt{SECURITY.md} files face in the vulnerability reporting process within open-source communities.
Specifically, we classified and analyzed the content of 711 randomly sampled issues related to \texttt{SECURITY.md}. We also conducted a quantitative comparative analysis of the close time and number of responses for issues concerning six community health files, including \texttt{SECURITY.md}. Our analysis revealed that 79.5\% of \texttt{SECURITY.md}-related issues were requests to add the file, and reports that included links were closed, with a median time that was 2 days shorter. These findings offer practical insights for improving security reporting policies and community management, ultimately contributing to a more secure open-source ecosystem.

\end{abstract}

%

\section{Introduction}
Most modern systems rely on Open Source Software (OSS), and a single security vulnerability can spread widely to multiple systems through webs of interconnected dependencies. However, if a developer who discovers a vulnerability reports it using public channels such as GitHub's Issues feature, the information runs the risk of being exploited by malicious third parties. Therefore, establishing a policy to properly and safely report vulnerabilities is crucial to reducing security risks for all projects that also rely on that software.

An effective solution for avoiding this risk is the introduction of a \texttt{SECURITY.md} file. This file was proposed at the GitHub Satellite held in May 2019 \cite{GitHubSatellite2019} and specifies secure reporting methods for vulnerabilities (e.g., specific email addresses or external forms) and information about supported versions. This allows vulnerability discoverers to report vulnerabilities directly to development teams without disclosing confidential information. GitHub has also published a \texttt{SECURITY.md} template in order to promote its use.

\begin{figure}[t]
  \centering
  \fbox{\includegraphics[width=0.9\textwidth]{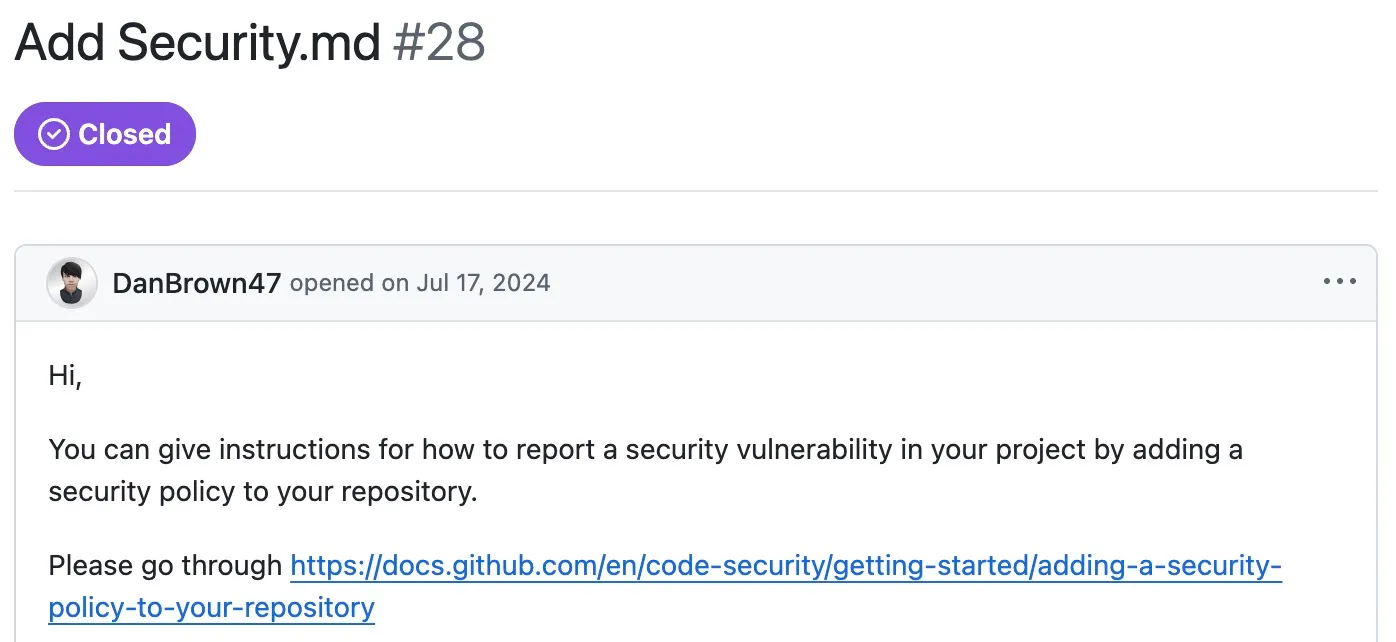}}
  \caption[a]{\texttt{SECURITY.md}-related issue\footnotemark}
  \label{fig:issue-example}
\end{figure}

Despite these recommendations, previous work found that the adoption rate of \texttt{SECURITY.md} files remains low at 7\%~\cite{DBLP:conf/icse/AyalaG23}, and its adoption has not yet become widespread.
The disconnect between GitHub's strong promotion of \texttt{SECURITY.md} and its limited adoption suggests there may be underlying barriers that prevent developers from implementing this security practice. Understanding these barriers is essential for developing strategies to improve adoption rates and ultimately enhance the security of the software supply chain. One valuable source of insight into these challenges is the discussions happening within the developer community itself—specifically, the GitHub issues where developers propose, debate, and implement \texttt{SECURITY.md} files. \autoref{fig:issue-example} illustrates an example of a common issue related to \texttt{SECURITY.md} files, where it is proposed that vulnerability reporting instructions be added to the project via a security policy.

In this study, we aim to clarify the problems developers may face in regards to \texttt{SECURITY.md} files, before and after adoption, through analysis of related issues. We analyzed 711 issues regarding \texttt{SECURITY.md} files and found that the most common request was for additions to \texttt{SECURITY.md}, accounting for 79.5\% of the total. Among these additional requests, issues that included detailed links related to \texttt{SECURITY.md}, such as documentation, took two days less to close. Our findings suggest an increased awareness of security policies by the community, and the effectiveness of providing example documentation.

\footnotetext{\url{https://github.com/incidentalhq/incidental/issues/28}}




\section{Related work}
Kancharoendee et al.\cite{DBLP:conf/saner/KancharoendeePJRKCRS25} investigated the relationship between the presence of a \texttt{SECURITY.md} file and a project's security score on GitHub. Their findings reveal that projects lacking a \texttt{SECURITY.md} file tend to have lower security scores from OpenSSF. Similarly, the work by Bühlmann et al.\cite{DBLP:conf/sac/BuhlmannG22} focused on how developers handle security issue reports on GitHub, noting an increasing trend in security-related issue reports and a decrease in their average resolution time.
These findings suggest that the adoption of \texttt{SECURITY.md} contributes to maintaining repository health and indicates a rising security awareness across the developer community. However, prior research~\cite{DBLP:conf/saner/KancharoendeePJRKCRS25,DBLP:conf/sac/BuhlmannG22} has not sufficiently addressed the factors hindering the adoption of \texttt{SECURITY.md} files or the specific challenges faced after implementation. Furthermore, while there has been research on analyzing issues for \texttt{README.md} files\cite{DBLP:journals/corr/abs-1802-08391}, there has been no research on issues related to \texttt{SECURITY.md}. Therefore, to bridge this gap in the literature, this study investigates and analyzes the challenges associated with \texttt{SECURITY.md} by focusing on the content of its related issues and the close time.


\section{Research Questions}
This study addresses the following two research questions (RQs).
\subsection*{RQ1: How long does it take to close \texttt{SECURITY.md} related issues?}
For this RQ, we investigate the close time of \texttt{SECURITY.md} related issues. Security-related issues are an area that requires early response, and so we ask how seriously developers take these issues. To answer this question, we compare:

\begin{itemize}
    \item[] \textbf{RQ1.1: Close time vs. other documentation}
    \item[] \textbf{RQ1.2: Close time with links}
    \item[] \textbf{RQ1.3: Number of responses to issues}
\end{itemize}

First, we evaluate how quickly issues related to \texttt{SECURITY.md} are addressed compared to other community health files (e.g., \texttt{CONTRIBUTING.md}, \texttt{SUPPORT.md}, etc.). Then we compare issues that include links providing detailed information about \texttt{SECURITY.md} with those that do not, and quantitatively evaluate whether the presence or absence of links to information about the installation of \texttt{SECURITY.md} has an effect on reducing the time required to close issues. Issues related to the addition of \texttt{SECURITY.md} need to be resolved more quickly. To provide more practical guidelines, it is necessary to analyze how differences in content affect resolution times, which is why this RQ was established. Finally, we quantitatively evaluate the level of community interest in security-related topics by comparing the ``number of responses,'' such as the number of comments and frequency of exchanges on an issue, between \texttt{SECURITY.md} and other files. Since this issue is related to security, we established this RQ based on the assumption that it would be discussed more than other files.

\subsection*{RQ2: What security policy related issues do developers create?}
This RQ focuses on the content of issues related to SECURITY.md.
In order to clarify the actual usage and issues of \texttt{SECURITY.md} in OSS, we investigated what kinds of issues are reported in GitHub Issues that mention \texttt{SECURITY.md}.
Examples include requests to add or modify \texttt{SECURITY.md}, but we also investigated what other issues there are and what percentage of them are included.


\section{Data Collection}
\label{sec:method}
In this study, we collected issues related to \texttt{SECURITY.md} or similar documents on GitHub. We utilized the GitHub REST API (v3) \cite{GitHubRESTAPI} to search public repositories for closed issues created between May 1, 2019 (i.e., the date GitHub introduced \texttt{SECURITY.md}), and June 30, 2025 (i.e., the date our data collection started). 
The search targeted issues that included any of the following documentation files in their titles or descriptions: 
\texttt{SECURITY.md}, 
\texttt{SUPPORT.md}, 
\texttt{LICENSE.md}, 
\texttt{GOVERNANCE.md}, 
\texttt{CONTRIBUTING.md}, 
and \texttt{CODE\_OF\_CONDUCT.md}. 
To accommodate API rate limits and ensure processing efficiency, data collection was continued until a total of 10,000 issues for each file type had been retrieved.

\medskip
\noindent
To ensure the quality of the collected issues, we applied the following filtering criteria:\vspace{-3mm}
\begin{itemize}
    \item Issues from repositories with fewer than 20 stars were excluded
    \item Non-English issues were filtered out using the \textit{langdetect} library to ensure consistent analysis.
    \item When multiple issues from the same repository were found, only the first instance was retained to avoid overrepresentation.
    \item Issues without a recorded close date were excluded to enable analysis of resolution times.
\end{itemize}

\noindent
After applying these filters, we obtained 15,192 issues. Table~\ref{tab:volume-of-issues} summarizes the number of issues collected. For each issue, we collected metadata including creation and closure dates to support our analysis of response times.

For the manual classification of \texttt{SECURITY.md} files (i.e., RQ2), we randomly sampled 711 issues from a total of 3,323.
Two authors independently reviewed and manually categorized each issue report according to its reporting purpose, employing an open card sorting methodology. The two inspectors
classified in the same way 97.5\% of the inspected issues,
with a Cohen’s kappa coefficient of 0.958, which demonstrates an
almost perfect agreement. The conflicting cases between the two inspectors were resolved through discussion, facilitated by a third author who acted as an adjudicator.

\begin{table}[t]
\centering
\caption{Number of issues related to each community health file.}
\label{tab:volume-of-issues}
\begin{tabular}{l r}
\toprule
 & \# of issues \\
\midrule
\texttt{GOVERNANCE.md}      & 240 \\
\texttt{SUPPORT.md}         & 478 \\
\texttt{LICENSE.md}         & 2,042 \\
\texttt{CODE\_OF\_CONDUCT.md} & 2,089 \\
\texttt{SECURITY.md}        & \textbf{3,323} \\
\texttt{CONTRIBUTING.md}    & 7,020 \\
\bottomrule
\end{tabular}
\end{table}


\section{Results}
\subsection*{RQ1.1: Do Developers Close SECURITY.md Related Issues Faster Than Other Documentation?}
\autoref{tab:close_time} summarizes the statistics of closing time  across different documentation file types. Issues associated with \texttt{SECURITY.md} exhibited a median close time of 7 days and a mean of 81.1 days, placing them among the relatively shorter durations observed. However, similar resolution times were found for other documentation types such as \texttt{LICENSE.md} and \texttt{CONTRIBUTING.md}, suggesting that the close time for \texttt{SECURITY.md} is not exceptionally brief in comparison.

In contrast, issues linked to \texttt{GOVERNANCE.md} stand out as a significant exception. These issues required substantially more time to resolve, with a median of 34 days and an average of 135.5 days. This finding implies that governance-related concerns may be inherently more complex or deprioritized in terms of timely resolution.

\begin{table}[th]
\centering
\caption{Days until close statistics for community health files.}
\label{tab:close_time}
\begin{tabular}{lrrrr}
\toprule
& {~~~~~Min.} & {~~~Median} & {~~~~~Mean} & {~~~~~Max.} \\
\midrule
\texttt{CODE\_OF\_CONDUCT.md} & 0 & 5 & 71.4 & \text{2,007} \\
\texttt{LICENSE.md} & 0 & 5 & 82.1 & \text{1,991} \\
\texttt{SECURITY.md} & \textbf{0} & \textbf{7} & \textbf{81.1} & \textbf{\text{1,977}} \\
\texttt{CONTRIBUTING.md} & 0 & 8 & 83.4 & \text{2,189} \\
\texttt{SUPPORT.md} & 0 & 11.5 & 112.2 & \text{2,007} \\
\texttt{GOVERNANCE.md} & 0 & 34 & 135.5 & \text{1,532} \\
\bottomrule
\end{tabular}
\end{table}

\subsection*{RQ1.2: Do Developers Close SECURITY.md Related Issues Faster When They Contain Links?}

We investigated whether the presence of additional information via links in the issue text influences the resolution time of issues. During data collection, two issues containing broken URLs were identified and excluded from the analysis because the broken links prevented accurate classification, thereby preserving the integrity and validity of our findings. 

Table~\ref{tab:with_without_link} presents the close-time statistics for issues with and without links. 
The median close time for issues without links was 9 days, whereas issues that included links were resolved in 7 days, 
suggesting that providing links enables maintainers to access relevant information more efficiently and to reduce communication overhead such as verification steps or follow-up inquiries. 
However, a formal comparison between issues with links ($n = 315$) and without links ($n = 97$) using the Mann--Whitney U test (two-sided, $\alpha = 0.05$) 
detected no statistically significant difference between the distributions ($U = 15{,}177.5$, $p = 0.9223$). This suggests that we need to investigate this research question deeper, such as using a bigger sample or investigating the different types of links.

\begin{table}[h]
\centering
\caption{Days until close statistics with link vs without link.}
\label{tab:with_without_link}
\begin{tabular}{lrrrrr}
\toprule
& \multicolumn{1}{l}{\# issues} & {~~~~~Min.} & {~~~Median} & {~~~~~Mean} & {~~~~~Max.}  \\
\midrule
With link & 315 & 0 & 7 & 82.9 & \text{1,309} \\
Without link & 97 & 0 & 9 & 66.4 & 764 \\
\bottomrule
\end{tabular}
\end{table}

\subsection*{RQ1.3: Do Developers Discuss SECURITY.md More Actively Than Other Documentation?}
\autoref{tab:close_time_with_response} shows the number of responses received for issues associated with various documentation files. Issues related to \texttt{SECURITY.md} had a median of 2 responses and an average of 3.1, indicating relatively active engagement. This level of interaction is comparable to LICENSE.md and \texttt{CODE\_OF\_CONDUCT.md}, which also had median values of 2 and slightly lower average responses. 

Notably, \texttt{GOVERNANCE.md} and \texttt{SUPPORT.md} issues received the highest average responses, at 4.6 and 4.7, respectively, suggesting that these topics may require more discussion or clarification. In contrast, \texttt{CONTRIBUTING.md} issues had the lowest median (i.e., 1) and average (i.e., 2.6) response counts, implying that contribution-related issues may be more straightforward or less debated. Overall, the data suggests that issues involving governance, support, and security tend to generate more community interaction, possibly due to their complexity or importance.

\begin{table}[th]
\centering
\caption{Number of issue responses for each file.}
\label{tab:close_time_with_response}
\begin{tabular}{lrrrr}
\toprule
& {~~~~~Min.} & {~~~Median} & {~~~~~Mean} & {~~~~~Max.} \\
\midrule
\texttt{SUPPORT.md} & 0 & 2 & 4.7 & 374 \\
\texttt{GOVERNANCE.md} & 0 & 2 & 4.6 & 131 \\
\texttt{SECURITY.md} & \textbf{0} & \textbf{2} & \textbf{3.1} & \textbf{374} \\
\texttt{CODE\_OF\_CONDUCT.md} & 0 & 2 & 2.9 & 357 \\
\texttt{LICENSE.md} & 0 & 2 & 2.8 & 77 \\
\texttt{CONTRIBUTING.md} & 0 & 1 & 2.6 & 85 \\
\bottomrule
\end{tabular}
\end{table}

\begin{tcolorbox}[
    breakable, 
    colback=gray!15, 
    colframe=black, 
    boxrule=0.5pt, 
    arc=0pt, 
    left=10pt, 
    right=10pt, 
    top=5pt, 
    bottom=5pt, 
    halign=left 
]
\textbf{Answer to RQ1:} Issues containing security-related URLs were closed two days earlier than those that did not, although no significant difference was observed in the overall close time or the number of responses for \texttt{SECURITY.md} issues compared to other document files.
\end{tcolorbox}

\begin{table}[t]
\centering
\small 
\caption{Classification results of SECURITY.md-related issues.}
\label{tab:classification_results}
\begin{tabular}{l @{\hspace{1em}} p{6cm} rr} 
\toprule
\textbf{Category} & \textbf{Description} & \textbf{~~\#} & \textbf{~~\%} \\
\midrule
\textbf{Creation Request} & Request to add a file, or report that a file does not exist. & 414 & 79.5 \\
\textbf{Revision Request} & Internal links, email revision requests, and content update requests. & 61 & 11.7 \\
\textbf{Reference} & Refer to Security.md or their contents. & 36 & 6.9 \\
\textbf{Educational Campaign} & Recommendation to use SECURITY.md. & 2 & 0.4 \\
\textbf{Other} & Requests to delete files, contributors' opinions, etc. & 8 & 1.5 \\
\bottomrule
\end{tabular}
\end{table}

\newpage
\subsection*{RQ2 What security policy related issues do developers create?}
We manually and independently classified the purposes of 711 issues randomly sampled from a total of 3,323 issues that contained the term \texttt{SECURITY.md}. \autoref{tab:classification_results} summarizes the distribution of issue purposes. During manual inspection, we excluded 190 issues that were not directly related to \texttt{SECURITY.md}. For example, those referencing directories or logs, or those automatically generated by bots such as \texttt{allstar-app} and \texttt{google-cloud-policy-bot}. After this filtering, 521 issues (73.3\%) remained for analysis.

The most prevalent category was ``Addition Requests,'' comprising 414 issues. Notably, 250 of these were submitted via the bug bounty platform \texttt{huntr.dev}~\cite{HuntrDev}, which specializes in OSS. On this platform, security researchers are rewarded for identifying and reporting vulnerabilities in OSS projects, thereby enhancing OSS security while incentivizing white-hat contributions.

The second most common category was ``Revisions'' (61 issues), which included requests to revise internal links, translate or update content, correct email addresses, relocate files for better accessibility, and fix content errors. Among the issues referencing \texttt{SECURITY.md}, 36 explicitly mentioned the file, with 25 directly referring to it---often with phrases like ``\textit{Please see \texttt{SECURITY.md} for details.}'' Of these, six discussed the file’s contents, such as version information or usage instructions, five raised questions about how to report vulnerabilities, indicating confusion about the appropriate reporting channel. Some GitHub users pointed out that there is also a reporting method called ``Report a security vulnerability,'' and it was unclear which reporting channel should be used.
Two issues served educational purposes, such as promoting best practices for community health files, discouraging the use of issue trackers for vulnerability disclosure, and advocating for the adoption of \texttt{SECURITY.md}. 
Finally, eight issues were categorized as ``Other.'' These included discussions about the file’s directory location, suggestions to synchronize or delete some community health files.

\begin{tcolorbox}[
    breakable, 
    colback=gray!15, 
    colframe=black, 
    boxrule=0.5pt, 
    arc=0pt, 
    left=10pt, 
    right=10pt, 
    top=5pt, 
    bottom=5pt, 
    halign=left 
]

\textbf{Answer to RQ2:} The types of issues related to \texttt{SECURITY.md} were 79.5 \% add requests, followed by revision requests and issues describing \texttt{SECURITY.md}. 
\end{tcolorbox}


\section{Threats to Validity}
With respect to internal validity, our analysis of resolution times may be influenced by unobserved factors such as repository size, popularity, ownership (organization vs. individual), or ecosystem. In addition, by considering only closed issues, the dataset may be biased toward more easily resolvable cases. 
Regarding external validity, our dataset was restricted to English-language issues from public GitHub repositories with at least 20 stars. Consequently, the findings may not generalize to smaller or private projects, non-English-speaking communities, or other platforms such as GitLab.

\section{Conclusion}
This study investigated \texttt{SECURITY.md} files, which are crucial for outlining secure vulnerability reporting procedures in open-source projects. We studied 711 randomly sampled issues related to \texttt{SECURITY.md}, finding that the most common type, at 79.5\% of requests, was to add the file. A significant portion of these addition requests, 48.0\%, originated from huntr.dev, an OSS-specialized bug bounty platform, while only 11.7\% were requests for revisions. This distribution indicates that \texttt{SECURITY.md} is still in its diffusion stage, not yet widely adopted.
Additionally, our findings indicate that while \texttt{SECURITY.md} helps clarify matters for developers, it can also increase confusion and burden.


\begin{credits}
\subsubsection{\ackname} 
We gratefully acknowledge the financial support of JSPS KAKENHI grants
(JP24K02921, JP25K21359), as well as JST PRESTO grant (JPMJPR22P3),
ASPIRE grant (JPMJAP2415), and AIP Accelerated Program (JPMJCR25U7).


\end{credits}

\bibliographystyle{IEEEtran} 
\bibliography{references}    

\end{document}